\documentclass{article}
\usepackage{arxiv}
\usepackage[utf8]{inputenc} 
\usepackage[T1]{fontenc}    
\usepackage{hyperref}       
\usepackage{url}            
\usepackage{booktabs}       
\usepackage{amsfonts}       
\usepackage{nicefrac}       
\usepackage{microtype}      
\usepackage{lipsum}		
\usepackage{graphicx}
\usepackage{doi}
\usepackage{cite}
\usepackage{amsmath,amssymb,amsfonts}
\usepackage{algorithmic}
\usepackage{amsmath}
\usepackage{textcomp}
\usepackage{multirow}
\usepackage{wrapfig}
\usepackage{tabularx}
\usepackage{ragged2e}
\usepackage{colortbl}
\usepackage{xcolor}
\usepackage{hyperref}

\usepackage{pifont}
\usepackage[super]{nth}
\usepackage{soul}
\usepackage[numbers]{natbib}

\usepackage{filecontents} 

\title{LOCAL: Low-Complex Mapping Algorithm for Spatial DNN Accelerators}

\author
 { 
{
Midia Reshadi}\thanks{This paper is published in: 2021 IEEE Nordic Circuits and Systems Conference (NorCAS), with DOI: 10.1109/NorCAS53631.2021.9599862.} \\
	School of Computer Science and Statistics\\
	Lero, Trinity College Dublin\\
	Dublin 2, Ireland \\
	\texttt{Midia.Reshadi@tcd.ie} \\
	\And
	{
	\hspace{1mm}David.Gregg} \\
	School of Computer Science and Statistics\\
	Lero, Trinity College Dublin\\
	Dublin 2, Ireland \\
	\texttt{David.Gregg@tcd.ie} \\
}

\hypersetup{
pdftitle={A template for the arxiv style},
pdfsubject={q-bio.NC, q-bio.QM},
pdfauthor={David S.~Hippocampus, Elias D.~Striatum},
pdfkeywords={First keyword, Second keyword, More},
}

\begin{document}
\maketitle

\begin{abstract}
Deep neural networks are a promising solution for applications that solve problems based on learning data sets. DNN accelerators solve the processing bottleneck as a domain-specific processor. Like other hardware solutions, there must be exact compatibility between the accelerator and other software components, especially the compiler. This paper presents a LOCAL (\underline{Lo}w \underline{C}omplexity mapping \underline{Al}gorithm) that is favorable to use at the compiler level to perform mapping operations in one pass with low computation time and energy consumption. 
We first introduce a formal definition of the design space in order to define the problem's scope, and then we describe the concept of the LOCAL algorithm.
The simulation results show 2$\times$ to 38$\times$ improvements in execution time with lower energy consumption compared to previous proposed dataflow mechanisms.
\end{abstract}

\keywords{Deep neural network \and spatial DNN accelerators \and mapping \and low-complex algorithm.}

\section{Introduction}
The growing amount of data brought learning-based systems from dreams to reality \citep{lecun2015deep}. Among machine learning methods, deep neural networks achieved better results; therefore, they have attracted much attention from both research and industry \citep{chen2016eyeriss},\citep{jouppi2017datacenter},\citep{NVDLA}. Deep neural networks are widely used today in many applications such as self driving cars \citep{bojarski2016end}, recommender systems \citep{karatzoglou2017deep}, and language translation \citep{wu2016google}. Due to performance bottleneck and high energy consumption of processing all parts of
DNN at the software level, domain-specific hardware, also called DNN accelerators, has been proposed in both resource constraints devices such as IoT edge \citep{ascia2019networks} and high performance cloud servers \citep{jouppi2017datacenter}. 

\begin{figure}[]
\centering
\centerline{\includegraphics[scale=0.70]{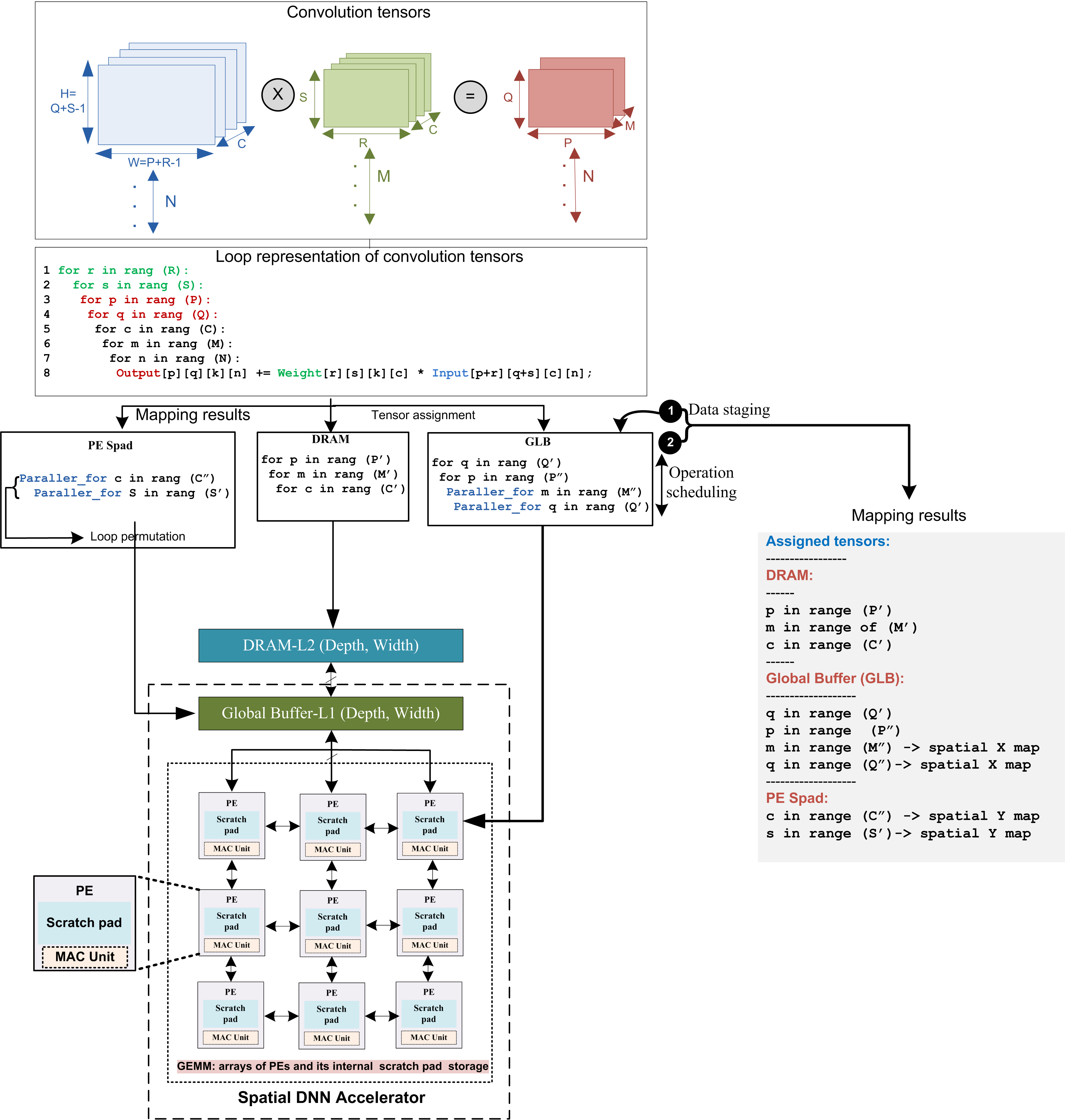}}
\caption{The overall concept of convolution tensors, their loop representation, and the mapping results of assigned tensors to memory elements including DRAM, GLB, and PE Spad of spatial DNN accelerator.}
\label{fig.1}
\end{figure}

The design space of deep neural network accelerators comprises hardware resource and data mapping strategy \citep{chen2016eyeriss}. The hardware resources generally consist of multi-level storage hierarchy to exploit data reuse \citep{chen2020noc} and an array of processing elements (PE) to perform parallel computations. Each processing element, as shown in Fig. \ref{fig.1}, comprises multiply and accumulate (MAC) logic connecting to local scratch pad memory, and all PEs are connected through the network-on-chip (NoC) interconnection.

Most of the proposed reference architectures \citep{chen2016eyeriss}\citep{chen2019eyeriss}\citep{chen2014diannao}\citep{shao2019simba} are different in their NoC topology and PE to memory connection. For instance, MAERI \citep{kwon2018maeri} is based on two binary fat trees, and Eyeriss \citep{chen2016eyeriss} and Google TPU \citep{jouppi2017datacenter} use 2D grid-style topology. In some accelerators, the PE array is connected to only one internal scratchpad memory and, in others, to multiple memory banks.

Another point in the DNN accelerator's design is the data mapping strategy that deals with two important tasks: \ding{182} Staging data on the HW resources and, \ding{183} operation scheduling (Fig. \ref{fig.1}). In other words, data mapping answers two fundamental questions, where data be placed? And when are operations performed? Fig. \ref{fig.1} shows the concept of the mapping on spatial DNN accelerator.

The basic mapping methods \citep{chen2016eyeriss} called \textit{dataflow} \citep{chen2017using} are based on reusing one of the primary convolution tensors, such as filter weights, input, or output activations. For instance, \textit{NVDLA} \citep{NVDLA} employs a \textit{weight stationary} strategy to reuse weights, \textit{Eyeriss} \citep{chen2016eyeriss} uses \textit{row stationary}, and \textit{ShiDianNao} \citep{du2015shidiannao} employs \textit{output stationary}.

Mapping is a more advanced concept than dataflow. It is about reusing multiple input values with specific ranges based on the shapes of DNN, dimensions of PE-arrays, levels and size of storage elements. Recently proposed mapping methods \citep{zhao2019mrna}, \citep{chatarasi2020marvel} have shown that, beyond the reuse of only one parameter and also correct placement of several parameters, more energy efficiency will be gained compared with conventional dataflow methods.

However, the vital issue in mapping is the vast search space of the solutions and complexity as far as we face $O(10 ^8)$ cases only in the fifth layer of VGG02 on Eyeriss accelerator that needs about $48$ hours to find an optimal mapping based on an exhaustive brute-force search method. This number of mapping choices can rise to $O(10^{72})$ for the 52-layer MobileNet-V2 \citep{kao2020confuciux}\citep{kao2020gamma}.

Typically, the mapping strategy is deployed in the design-phase \citep{mirmahaleh2019flow}\citep{mirmahaleh2020flow} or compile-time \citep{parchamdar2020data}. Hence, finding the optimal mapping in the \textit{design-time} is essential but critical when it needs to be solved in \textit{compile-time}. Several mapping algorithms have been proposed using evolutionary or learning-based algorithms to solve a problem \citep{kao2020confuciux}\citep{kao2020gamma}, although, applying them at the compiler level extends the compile time due to many running iterations.

To tackle the complexity and reducing mapping time, we propose a mapping algorithm called \textit{LOCAL}. LOCAL's primary goal is finding a close-to optimal mapping in low computation time with admissible energy consumption.

We formally describe the problem to reach a clear and structured solution then we present the LOCAL mapping algorithm. Finally, we contrast our approach against other dataflow mechanisms, including weight, output, and row stationary in NVDLA, ShiDianNao, and Eyeriss accelerators, respectively.

Therefore, the main contributions of this paper are as follows:

\begin{itemize}
\item \textbf{Problem formulation.} We formulate the problem to achieve an accurate definition of the scope of the problem. Problem formulation defines the scope and dimension of the problem, so that helps to find an optimal solution between massive solution options. 

\item \textbf{A low-complexity one-pass mapping algorithm.} We propose a low complexity mapping algorithm called LOCAL with 2$\times$ to 38$\times$ higher speed compared to other dataflow mechanisms.

\item \textbf{Usability at the compiler level.} LOCAL can perform a fast mapping at runtime in a single pass, making it a favorable option for compiler-level implementation.

\item  \textbf{Low energy consumption.} The second goal of LOCAL is reducing energy consumption. Simulation results show LOCAL is more energy-efficient than conventional dataflow techniques.
\end{itemize}

The paper is organized as follows: Section II provides preliminaries on convolution and spatial DNN accelerator; Section III presents the paper's motivation; Section IV presents problem formulation; Section V describes LOCAL mapping algorithm in detail; Section VI presents comprehensive evaluations; Section VII presents related work and Section VIII concludes the idea.

\section{Preliminaries}
The main assumptions of this paper are DNN and accelerator, and the final goal is the mapping algorithm that accepts those assumptions as inputs. This section introduces the basic definitions of convolution in modern DNNs and the structure of spatial DNN accelerators.

\subsection{Convolutions}
\textit{Definition 2.1:} A convolution tensor (CT) comprises three tensors: filter weights, input and output feature maps. Hence, we define CT as:
\begin{equation}
CT = \{Weight, Input, Output\}
\end{equation}
which
\begin{equation}
CT\in \mathbb{R}^{dimension}    
\end{equation}
and, dimensions are:
\begin{equation}
dimension=\{N,M,C,R,S,W,H,P,Q\}    
\end{equation}
which, $W\in \mathbb{R}^{MCRS}$, $I\in \mathbb{R}^{NCHW}$, and $O\in \mathbb{R}^{NMPQ}$, are filter weights, input and output feature maps, respectively.
\begin{equation}
ct_i\in CT    
\end{equation}
\begin{equation}
CT=\{ct_1,ct_2,ct_3\}    
\end{equation}
\begin{equation}
CT=\{\mathbb{R}^{MCRS},\mathbb{R}^{NCHW},\mathbb{R}^{NMPQ}\}    
\end{equation}
We specify a shape of tensor as $r^x\in tc_i$ that is:
\begin{equation}
r^x\in \mathbb{R}^y \;\;\;\;\;\; x,y\in dimension    
\end{equation}
For example:
\begin{equation}
r^{c}\in Weight\rightarrow r^c\in \mathbb{R}^{MCRS}    
\end{equation}

We may summarize that, various CNNs have different tensor dimensions. In the software layer each tensor is defined as a loop-nests; for example, the related loop of $r^c$ is:
\begin{equation}
r^{c}\rightarrow for\;\;c\;\;in\;\;range (C)    
\end{equation}

For better illustration, Fig. \ref{fig.1} depicts the loop representation and matrix shape of convolution tensors. The mapping result is also shown in Fig. \ref{fig.1} which are mapped tensors to the storage elements of an accelerator.

\subsection{Spatial DNN accelerator}
\textit{Definition 2.2:} A spatial DNN accelerator (SPA) comprises an array of multi-level storage elements with hierarchy and array of processing elements:
\begin{equation}
SPA=\{Storage[i,j,k], PE[m,n]\}    
\end{equation}
Storage elements are: 
\begin{equation}
s_{i_{j,k}}\; , \;\;\;\ S\in \mathbb{R}^{i,j,k}    
\end{equation}

\begin{figure}[]
\centering
\centerline{\includegraphics[scale=0.57]{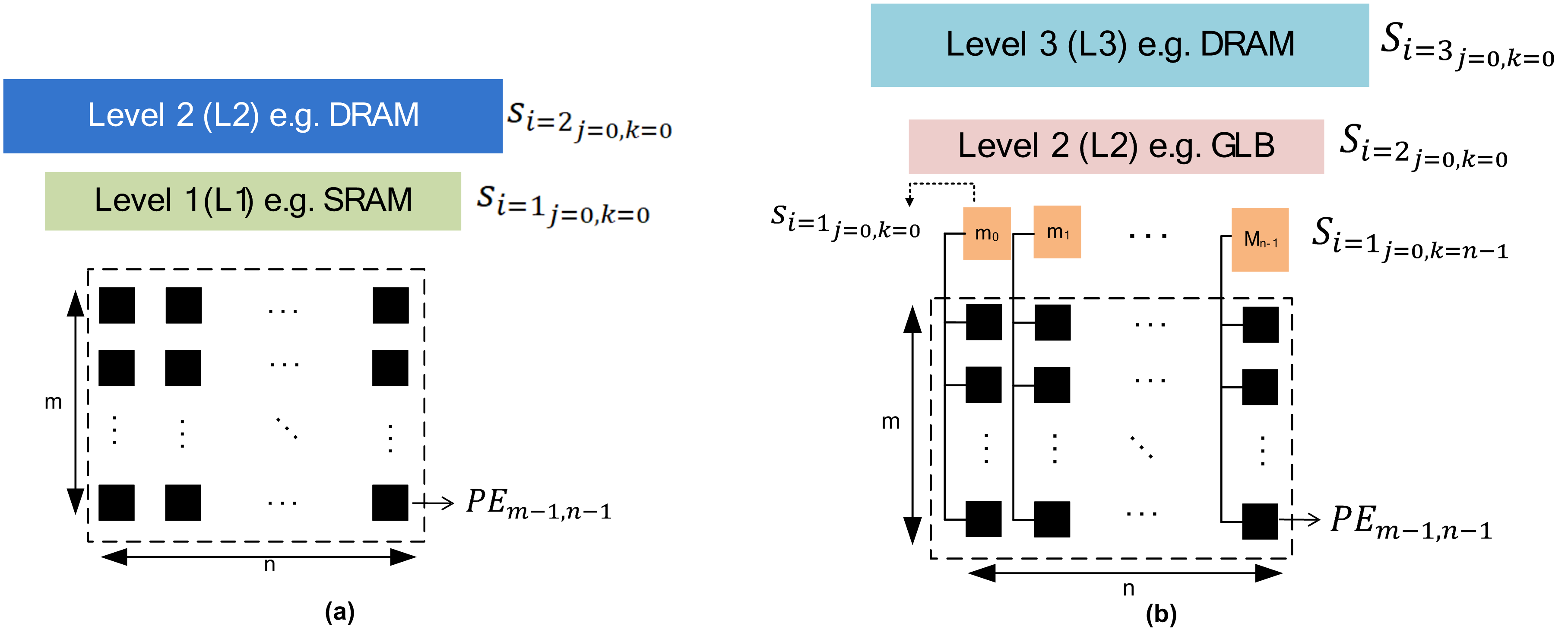}}
\caption{(a) \textit{NVDLA-style}: 1D memory array and a 2D array of processing elements (PE) in $(m,n)$ dimensions. (b) \textit{Eyeriss-style}: 1D memory in each level, multiple memory elements in level 1 $(1,n)$ , and 2D arrays of PEs in $(m,n)$.}
\label{fig.2}
\end{figure}

That $i$ means the hierarchy level, and $j$ and $k$ are the indices of memory elements in a 2D space.
Fig. \ref{fig.2}(b), depicts an example of representing a memory element as $s_{{i=1}_{j=0,k=n-1}}$, that means the position of the memory element $m_{n-1}$ in the $1^{st}$ level of hierarchy in $j=0$ and $k=n-1$ dimension. For better readability, we present $s[i,j,k]$ as $s_{i_{j,k}}$.

The size of each storage element $s\in S$ at the $i^{th}$ level is equal to: 
\begin{equation}
\left| s \right|=Depth\times Width    
\end{equation}

Meanwhile, the array of processing elements is defined as: 
\begin{equation}
PE_{m,n}\in \mathbb{R}^{x,y}    
\end{equation}

which, all $(m\times n)$ PEs are connected by a Network-on-Chip (NoC). 

Based on the connection between global memory and a array of processing elements, we consider two different accelerator types.

The first architecture, called \textit{NVLDA-style}, comprises two internal memory elements ($L0$ registers at PE and global buffer as $L1$). The second architecture, called \textit{Eyeriss-style}, includes three internal memory levels, including $L0$ at PE, 1D array of multiple elements as $L1$, and global buffer as $L2$. The purpose of using $L1$ memory elements is to connect each $m_k$ elements with the range of $k=0\to n-1$ to the column of $PE_{i,j}$ in the range of $i=0\to m-1$ and $j=0\to n-1$. Therefore, we can summarize the memory to PE connections as follows:\\
\textit{NVDLA-style} (Fig. \ref{fig.2}(a)):

Level 1 ($L1$): $s_{i=1_{j=0,k=0}}$
\begin{equation}
s_{i=1_{j=o,k=0}}\;connects\;to\;PE_{i=0 \to m-1,j=0\to n-1}
\end{equation}
\textit{Eyeriss-style} (Fig. \ref{fig.2}(b)):

Level 1 (L1): $s_{i=1_{j=0,k=0 \to (m-1)}}$
\begin{equation}
s_{{i=1}_{j=0,k=0}}\;\;connects\;to\;PE_{m=0\to (m-1),n=0}    
\end{equation}
\begin{equation}
s_{{i=1}_{j=0,k=n-1}}\;\;connects\;to\;PE_{m=0\to (m-1),n=(n-1)}    
\end{equation}

Fig. \ref{fig.2} shows the summarized and abstract structure of the spatial DNN accelerator. We represent the structural features of Eyeriss and NVDLA-style accelerators to achieve a clear problem definition.

\subsection{Mapping Algorithm}
The mapping of the convolution tensors (CT) onto the spatial DNN accelerator (SPA) is defined by the \textit{mapping} function, including the following operations:
\begin{enumerate}
    \item \textbf{\textit{Assignment.}} Tensor assignment means assigning tensors to storage elements: 
\begin{equation}
ct_i\in CT\;assign\;to\;s_{i_{j,k}}\in S
\end{equation}
\item \textbf{\textit{Bounding.}} Bounding is the process of limiting tensor dimensions so that they are smaller than the size of storage elements:
\begin{equation}
\left| CT \right|\le \left| S \right|   
\end{equation}

Hence, the tensor assignment is bounded by a specific range:
\begin{equation}
ct_i[0,range]\in CT\;assign\;to\;s_{i_{j,k}}\in S
\end{equation}

For example $r^Q$ tensor with the following for-loop representation is assigned to $s_{{i=1_{j=0,k=0}}}$ memory element at the $1^{st}$ level of memory hierarchy:
\begin{align*}
for\ Q\ in\ [0,5)\ assigned\ to\ s_{i=1_{j=0,k=0}}
\end{align*}
\item \textbf{\textit{Scheduling.}} Tensor order scheduling is the permutation of every allocated tensor. To be more precise, it involves defining the tensor order.

\begin{equation}
ct_i,\ ct_j,\ ct_k\ \in CT\ assign\ to\ L_i    
\end{equation}
such that:
\begin{align*}
\ \ ct_i[0,rang_i)\ \ \\
\ ct_j[0,rang_j)\ \\
ct_k[0,rang_k) 
\end{align*}
Thus, the permutation is in the order of $ct_i$,$ct_j$,$ct_k$, respectively. For example, at the $i^{th}$ level of memory, we may have the following order:
\begin{align*}
\ \ for\ M\ in\ [0,5)\ \ \\
\ for\ Q\ in\ [0,4)\ \\
for\ P\ in\ [0,6)    
\end{align*}

\item \textbf{\textit{Parallelization.}} Spatial partitioning of assigned tensors is known as parallelization. More specifically, it refers to assigning a tensor to a subset of PEs in order to carry out parallel computation.
\begin{equation}
Spatial\ computing\to ct_i[0,rang)
\end{equation}
which, $ct_i[0,rang)$ is:
\begin{equation}
ct_i[0,rang)\in CT\ on\ PE_{[i\ to\ j]\ (x|y)} \in PEs
\end{equation}
For example:

\textit{Parallel\_for\ S\ in\ [0,7)\ on\ PE[0-7)\ Spatial\ X\ dimension}

Fig. \ref{fig.1} shows the example of final mapping results includes assigned, bounded, scheduled, and parallelized tenors to storage elements (DRAM, GLB, and PE Spad).    
\end{enumerate}

\section{Motivation}
\begin{figure}[]
\centering
\centerline{\includegraphics[scale=0.38]{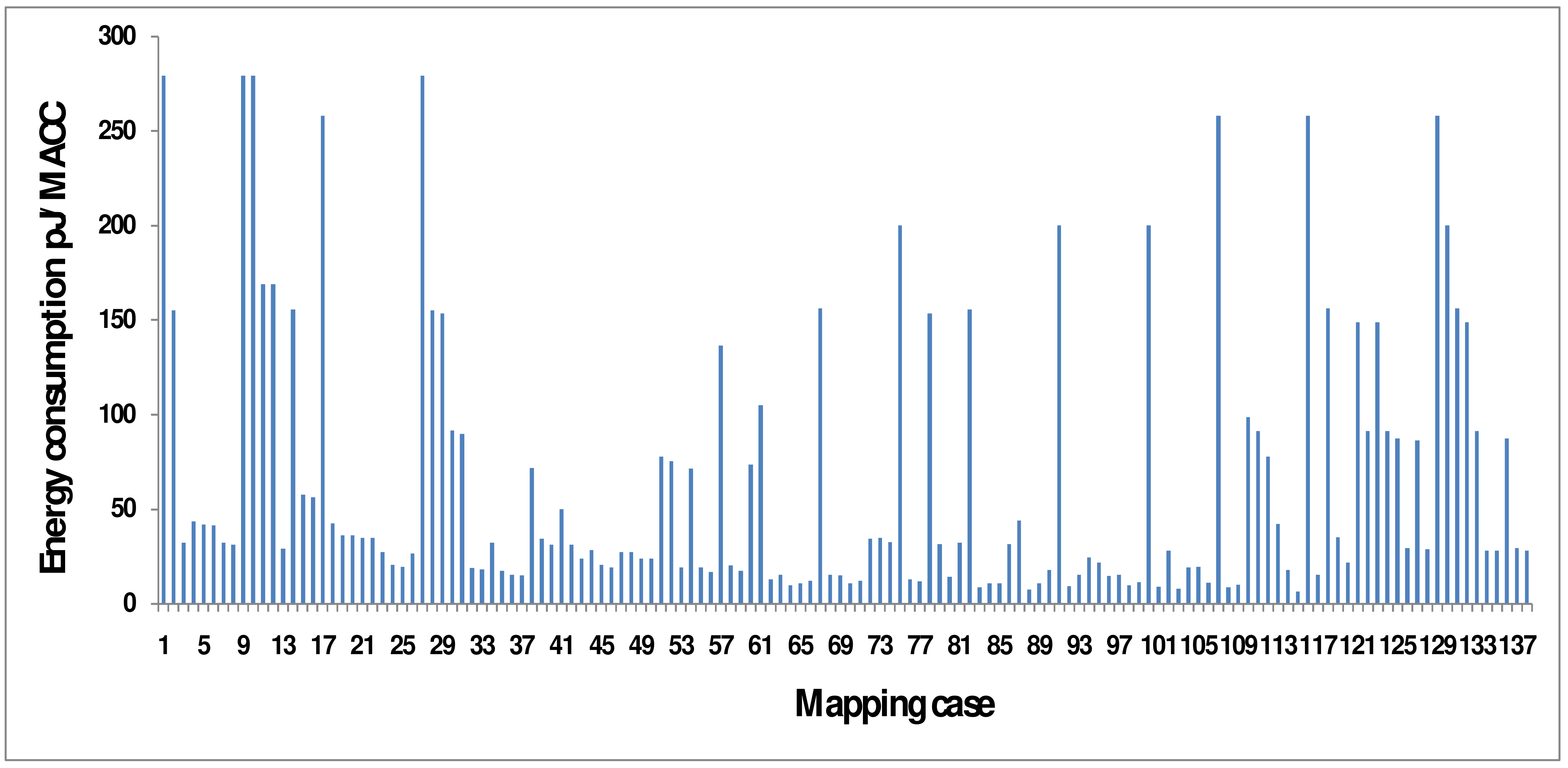}}
\caption{Energy consumption of random mapping.}
\label{fig.3}
\end{figure}

To show the importance of the mapping strategy on energy consumption, we conducted an experiment generating 3,000
random mapping cases without any heuristics. We randomly map the fifth layer of \textit{VGG02} on the Eyeriss-style accelerator according to the configuration shown in Table 1. The results are classified into three categories, including \textit{random\_max}, \textit{random\_med}, and \textit{random\_min}, as the cases with maximum, median, and minimum energy consumption, respectively. Fig. \ref{fig.3} shows, there is 77\% difference between the \textit{random\_max} and the \textit{random\_med} and 90\% between the \textit{random\_med} and \textit{random\_min} cases. As we can see, the random mapping does not necessarily find the optimal solution, but it still manages to save almost 90\% when compared to the median and minimum solutions. The energy-saving rises when the number of randomly generated mapping increases.

\begin{table}
\centering
\caption{The features of Eyeriss spatial DNN accelerator (SPA) and the shapes of the fifth layer of VGG02 convolution tensors (CT).}
\begin{tabular}{|c|c|l|c|c|} 
\cline{1-2}\cline{4-5}
\multicolumn{2}{|c|}{\textbf{SPA architecture}}                                         &  & \multicolumn{2}{c|}{\textbf{CT shapes }}                       \\ 
\cline{1-2}\cline{4-5}
SPA                                                              & Eyeriss              &  & CT & \begin{tabular}[c]{@{}c@{}}Layer 5\\VGG\_02\end{tabular}  \\ 
\cline{1-2}\cline{4-5}
\begin{tabular}[c]{@{}c@{}}On-chip \\storage~levels\end{tabular} & 2                    &  & C  & 128                                                       \\ 
\cline{1-2}\cline{4-5}
DRAM(width)                                                      & 64                   &  & M  & 256                                                       \\ 
\cline{1-2}\cline{4-5}
$L_1$ (depth,width)                                           & (16384, 64)          &  & N  & 1                                                         \\ 
\cline{1-2}\cline{4-5}
$L_0$ (depth,width)                                           & (16,16)              &  & P  & 56                                                        \\ 
\cline{1-2}\cline{4-5}
PE array                                                         & (12,14)              &  & Q  & 56                                                        \\ 
\cline{1-2}\cline{4-5}
\multicolumn{1}{l}{}                                             & \multicolumn{1}{l}{} &  & R  & 3                                                         \\ 
\cline{4-5}
\multicolumn{1}{l}{}                                             & \multicolumn{1}{l}{} &  & S  & 3                                                         \\
\cline{4-5}
\end{tabular}
\end{table}

Based on CT shapes and levels of the memory hierarchy in Eyeriss, the map-space is $(n!)^m$, while $n$ is the number of loop-nests that can be swapped randomly, and $m$ is the number of storage levels. Since the mapping space of the fifth layer of \textit{VGG02} with six nested loop on Eyeriss with three storage levels is equal to $O(10^8)=(6!)^3$. As a result, calculating the energy and latency of all mapping cases with searching by exhaustive brute-force methods will take a long time.
In this example, the structure of an accelerator is already specified, but if we need to decide on the number of processing elements, we will have $O(10^9)=64^2\times 224^2\times 3^2$ design cases for the second layer of \textit{VGG16} $(K=64, C=64, Y=224, X=224, R=3, S=3)$.

Thus, the entire design space, including the accelerator configuration and mapping-space, is equal to $O(10^{17})=64^2\times 224^2\times 3^2 \times6!^3$. 

As we can see, this design space is hard to enumerate, so a low-complex mapping algorithm is needed. The following section introduces the \textit{LOCAL} algorithm that finds the nearly optimal energy-efficient solution $2\times$$-$$38\times$ faster than other dataflow mechanisms.

\section{Problem formulation}
Given the convolution tensor characteristics and the structure of the  spatial DNN accelerator, our objective is to map the convolution tensors (CT) to the spatial DNN accelerator (SPA) such that the energy consumption is minimized under a minimum complex mapping algorithm. Of note, maximizing the number of inferences per second causes maximizing the PE utilization and minimizing energy consumption \citep{sze2020evaluate}. More formally:

\textbf{Given:} A convolution tensors (CT), and a spatial DNN accelerator (SPA).

\textbf{Find:} a mapping function that maps convolution tensors to a spatial DNN accelerator to minimize:
\begin{equation}
min\ \{energy\}    
\end{equation}
and maximizing:
\begin{equation}
max\ \{PE\ utilization\}    
\end{equation}
such that:
\begin{equation}
utilization\ of\ PEs=\frac{number\ of\ active\ PEs}{number\ of\ PEs}    
\end{equation}

Accordingly, high PE utilization is achieved by keeping the maximum number of PEs active, and this is achieved by proper data assignment and operation scheduling.

\section{LOCAL mapping algorithm}
\begin{figure}[]
\centering
\centerline{\includegraphics[scale=0.74]{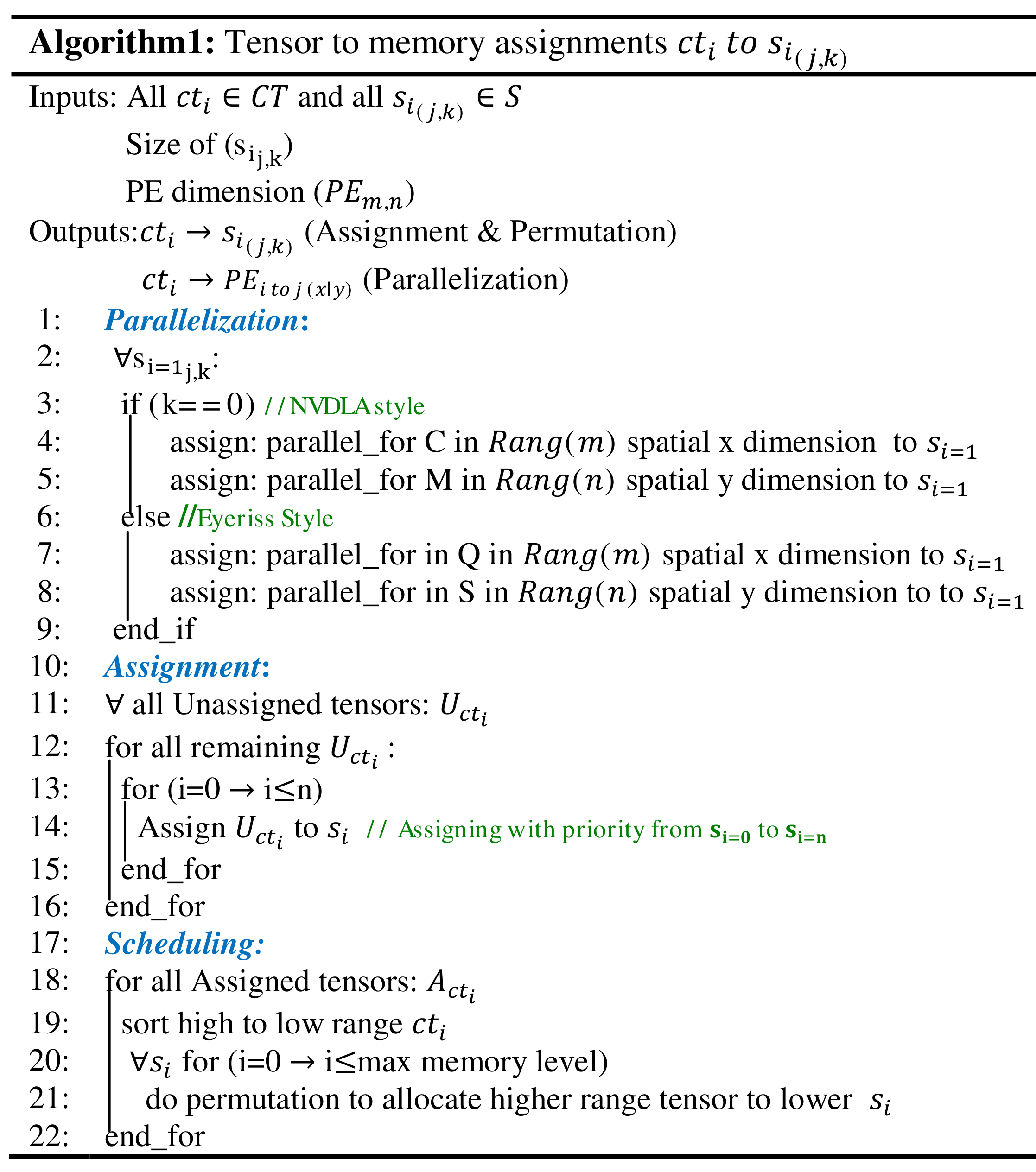}}
\caption{The pseudo code of \textit{LOCAL} algorithm.}
\label{fig.4}
\end{figure}

The main feature of the LOCAL mapping algorithm is low complexity, which provides an appropriate mapping in one pass and a short amount of time.
The mapping algorithms that were previously proposed \citep{kao2020confuciux}\citep{kao2020gamma} were iterative and had long execution times.
The main idea behind the LOCAL algorithm is to achieve maximum parallelism by performing spatial mapping of effective tensors, because high parallelism leads to increased PE utilization, as presented in Eq.(25).
As Fig. \ref{fig.4} shows, the LOCAL algorithm takes convolution tensors ($ct_i\in CT$), storage elements ($s_{i_{j,k}} \in S$), and dimensions of PEs as input and gives the final mapping as assigned tensors to storage elements with specified ranges, schedules, and parallelized features.

\begin{figure}[]
\centering
\centerline{\includegraphics[scale=0.85]{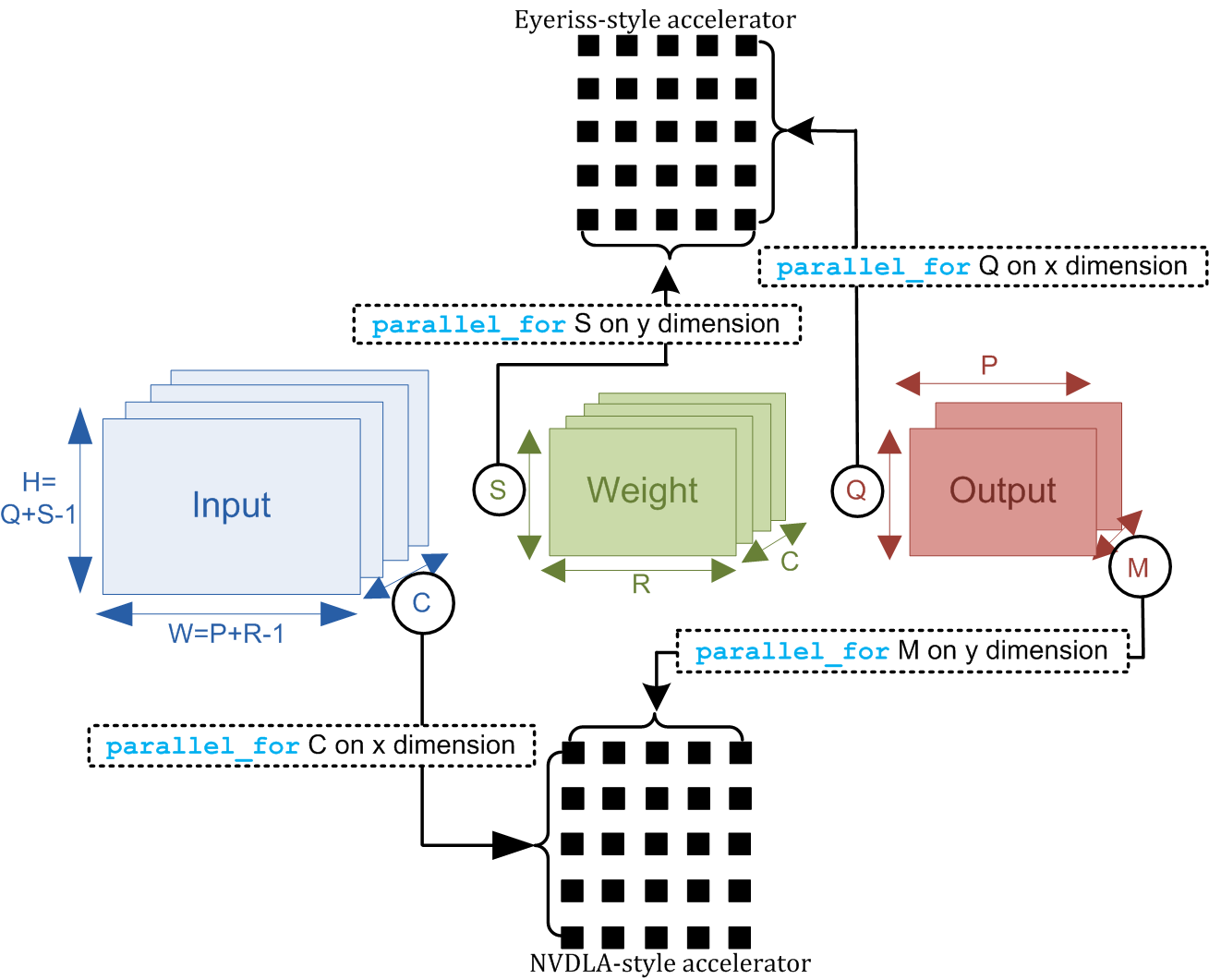}}
\caption{Tensor spatial mapping (\texttt{\textbf{parallel\_for}}) in Eyeriss and
NVDLA-like accelerators.}
\label{fig.5}
\end{figure}

According to Fig. \ref{fig.4}, the LOCAL mapping algorithm consists of three main steps, \textit{parallelization}, \textit{assignment}, and \textit{scheduling}.

Parallelization is the first step due to its importance. In this step, the parallelized tensors are considered based on the type of accelerator. According to Fig. \ref{fig.2}, Eq. (14-16), the main difference between Eyeriss and NVDLA style is the number of storage blocks at the level $i=1$, and also their connection to an array of PEs. While the type of accelerator is NVDLA-style, $C$ and $M$ as effective shapes are mapped spatially on the $x$ and $y$ dimensions of PE-array, respectively (lines 3-5), which causes increasing PE utilization as well as energy efficiency (Fig. \ref{fig.5}). Likewise, the two most effective shapes in the Eyeriss-style, $Q$ and $S$, are mapped in parallel to the $x$ and $y$ dimensions of PE-array, respectively (lines 7 and 8) (Fig. \ref{fig.5}). 

The next step, called an assignment, assigns the rest of the unassigned tensors to memory elements with priority from the lowest to the highest level (lines 11-16).

The final phase is scheduling, which permutes assigned tensors to give lower-level memory elements larger range due to the lower energy cost at lower levels of memory than at higher levels (Lines 18-22). After running the LOCAL algorithm, the assigned tensors with their range, parallelization, and loop-scheduling are determined to be the algorithm's output.

\section{Evaluation}
\subsection{Method}
\begin{figure}[]
\centering
\centerline{\includegraphics[scale=0.40]{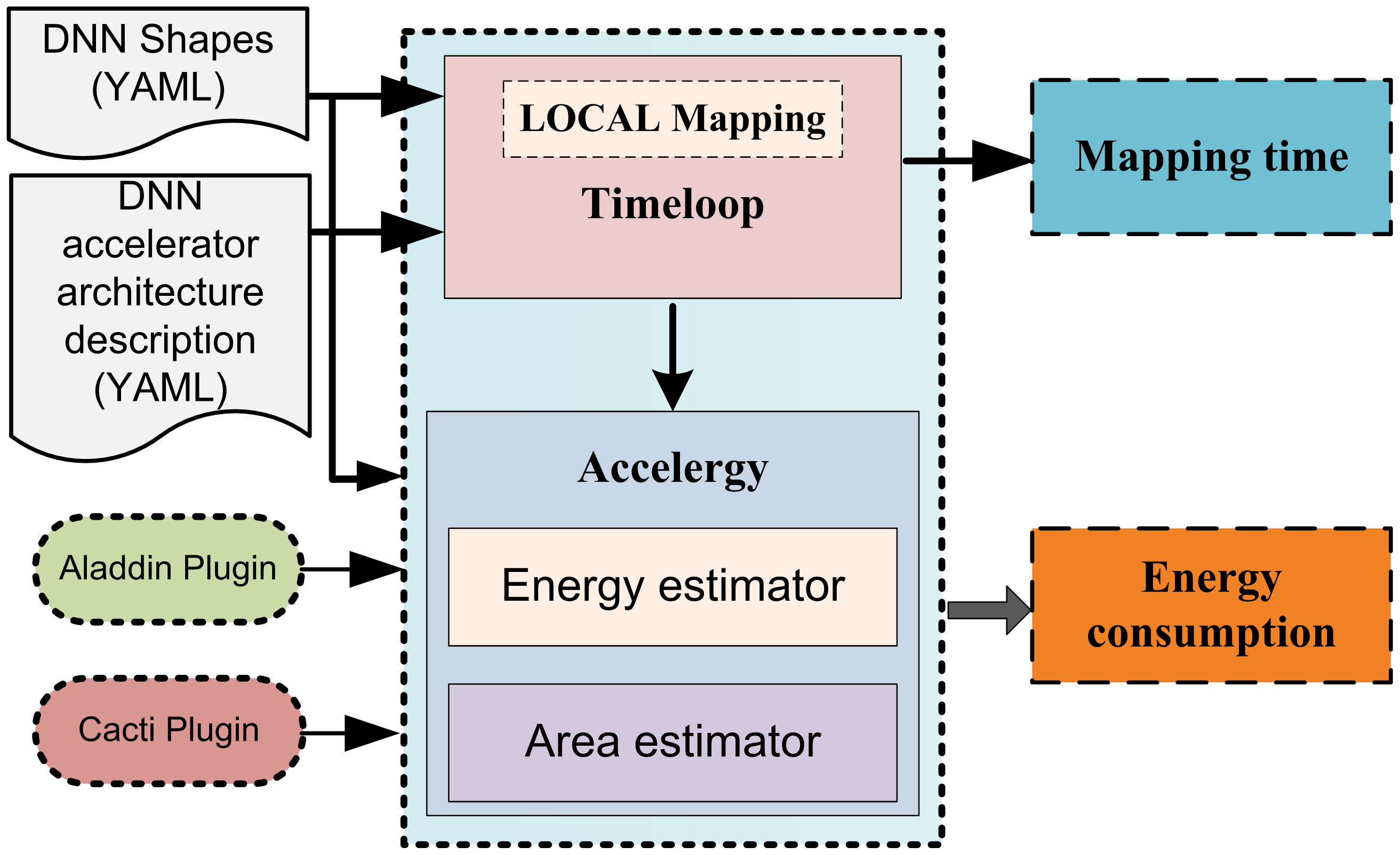}}
\caption{Simulation framework.}
\label{fig.6}
\end{figure}

To simulate the LOCAL algorithm and compare it with other dataflow mechanisms, we added the LOCAL mapping to the source code of  the Timeloop-Accelergy framework \citep{wu2019accelergy}\citep{parashar2019timeloop}\citep{LOCALMapping}. Fig. \ref{fig.6} shows the simulation framework.

\subsection{Simulation Workloads}
To achieve a fair simulation, we categorized the workloads based on the magnitude of four main parameters of the convolution layers, including $C$, $M$, $P$, and $Q$. This category, with more details about the number of MAC operations, is shown in Table 2. We applied our proposed mapping to the Eyeriss, NVDLA, and ShiDiaNao accelerators and compared it with the row, weight, and output stationary, respectively. 
Comparisons are made based on two main parameters: energy consumption and mapping time. Mapping time is based on seconds and is equal to the duration of time it takes to find the proper map. Table 3 shows the mapping times for Eyeriss, NVDLA, and ShiDianNao with LOCAL mapping and their dataflows. The calculation time of row, weight, and output stationary are extracted from the Timeloop-Accelery framework by defining data-reuse constraints. Furthermore, the mapping time of the LOCAL algorithm is evaluated based on employing the LOCAL mapping function inside the Timeloop. It should be noted that the different mapping times of the LOCAL algorithm in Table 3 are due to the variation of DNN layer shapes. 

\begin{table} 
\centering
\caption{Workload categories}
\begin{tabular}{c|l|c} 
\hline
\textbf{Category}                                                             & \multicolumn{1}{c|}{\textbf{Workload}} & \begin{tabular}[c]{@{}c@{}}\textbf{Number of MAC}\\\textbf{operations}\end{tabular}  \\ 
\hline
\multirow{3}{*}{High C value}                                                 & $22^{nd}$ conv layer of Resnet50              & 51380224                                                                             \\ 
\cline{2-3}
                                                                              & $23^{rd}$ conv layer of SqueezNet          & 5537792                                                                              \\ 
\cline{2-3}
                                                                              & $9^{th}$ conv layer of VGG16               & 1849688064                                                                           \\ 
\hline
\multirow{3}{*}{High M value}                                                 & $25^{th}$ conv layer of SqueezNet          & 24920064                                                                             \\ 
\cline{2-3}
                                                                              & $24^{th}$ conv layer of ResNet50           & 51380224                                                                             \\ 
\cline{2-3}
                                                                              & $8^{th}$ conv layer of VGG16               & 924844032                                                                            \\ 
\hline
\multirow{3}{*}{\begin{tabular}[c]{@{}c@{}}High P and Q\\values\end{tabular}} & $1^{st}$ conv layer of SqueezNet           & 708083712                                                                            \\ 
\cline{2-3}
                                                                              & $1^{st}$ conv layer of ResNet50            & 472055808                                                                            \\ 
\cline{2-3}
                                                                              & $1^{st}$ conv layer of VGG16               & 86704128                                                                             \\
\hline
\end{tabular}
\end{table} 

As can be seen from Tables 3, LOCAL reaches $34\times$, $38\times$, and $49\times$, faster calculation time than row, output, and weight stationaries. Since in those dataflows, we still need many comparisons to select the appropriate case despite the definition of numerous constraints. It is noteworthy that we have several cases in each stationary method that differ due to parallel for-loops. For this reason, we need many comparisons in choosing the proper case.

\begin{figure}[]
\centering
\centerline{\includegraphics[scale=0.6658]{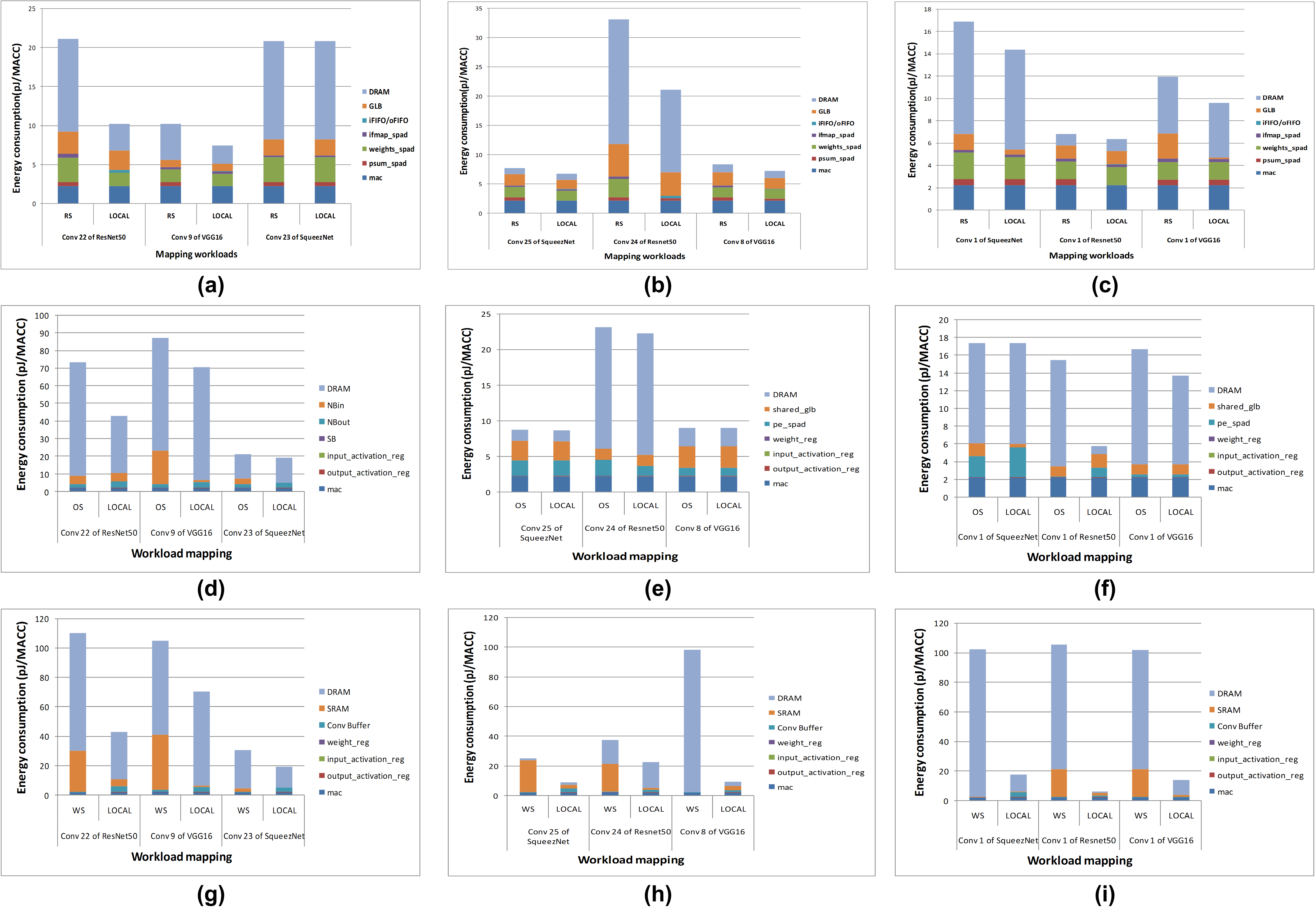}}
\caption{The energy consumption of LOCAL mapping algorithm comparing with row stationary dataflow in Eyeriss, output stationary dataflow in Shi-diannao, and weight stationary in NVDLA. (a) Energy consumption of row stationary and LOCAL mapping in Eyeriss with High $C$ value workload. (b) Energy consumption of row stationary and LOCAL mapping in Eyeriss with High $M$ value workload. (c) Energy consumption of row stationary and LOCAL mapping in Eyeriss with High $P$ and $Q$ values workload. (d) Energy consumption of output stationary and LOCAL mapping in Shi-diannao with High $C$ value workload. (e) Energy consumption of output stationary and LOCAL mapping in Shi-diannao with High $M$ value workload. (f) Energy consumption of output stationary and LOCAL mapping in Shi-diannao with High $P$ and $Q$ values workload. (g) Energy consumption of weight stationary and LOCAL mapping in NVDLA with High $C$ values workload. (h) Energy consumption of weight stationary and LOCAL mapping in NVDLA with High $M$ values workload. (i) Energy consumption of weight stationary and LOCAL mapping in NVDLA with High $P$ and $Q$ values workload.}
\label{fig.7}
\end{figure}

Fig. \ref{fig.7} shows the energy consumption of previously proposed dataflows including, row, output, and weight stationary for Eyeriss, ShiDianNao, and NVDLA accelerators. As expected, a large portion of the energy consumption is related to DRAM, and we can also conclude that if the amount of data movement between memory elements and the array of processing elements is reduced, we can achieve higher energy efficiency. Another conclusion that can be drawn from the results is that the LOCAL algorithm has achieved acceptable results in terms of energy consumption in a short processing time compared to other dataflows.

\begin{table*} 
\centering
\caption{The mapping time of Eyeriss, NVDLA, and ShiDianNao with LOCAL mapping and their dataflows.}
\begin{tabular}{|>{\centering\arraybackslash}m{1.3cm}|>{\centering\arraybackslash}m{2.78cm}|>{\centering\arraybackslash}m{1.5cm}|>{\centering\arraybackslash}m{1.5cm}|>{\centering\arraybackslash}m{1.5cm}|>{\centering\arraybackslash}m{1.5cm}|>{\centering\arraybackslash}m{1.5cm}|>{\centering\arraybackslash}m{1.5cm}|}

\hline
\rowcolor[rgb]{0.753,0.753,0.753} \multicolumn{1}{|l|}{\textbf{Workload}}                                                            & \textbf{\textit{Convolution}}       & \textbf{\textit{Mapping mechanism}} & \textbf{\textit{Mapping time (sec)}} & \textbf{\textit{Mapping mechanism}} & \textbf{\textit{Mapping time (sec)}} & \textbf{\textit{Mapping mechanism}} & \textbf{\textit{Mapping time (sec)}}  \\ 
\hline
\multirow{6}{*}{\begin{tabular}[c]{@{}c@{}}\textbf{\textit{High C}}\\\textbf{\textit{value}}\end{tabular}}                           & \multirow{2}{*}{Resnet50: Conv 22}  & RS                                  & 87                                   & OS                                  & 576                                  & WS                                  & 127                                   \\ 
\cline{3-8}
                                                                                                                                     &                                     & LOCAL                               & 16.2                                 & LOCAL                               & 15                                   & LOCAL                               & 6                                     \\ 
\cline{2-8}
                                                                                                                                     & \multirow{2}{*}{VGG16: Conv 9}      & RS                                  & 170                                  & OS                                  & 137                                  & WS                                  & 68                                    \\ 
\cline{3-8}
                                                                                                                                     &                                     & LOCAL                               & 10                                   & LOCAL                               & 15                                   & LOCAL                               & 9                                     \\ 
\cline{2-8}
                                                                                                                                     & \multirow{2}{*}{SqueezNet: Conv 23} & RS                                  & 17                                   & OS                                  & 125                                  & WS                                  & 21                                    \\ 
\cline{3-8}
                                                                                                                                     &                                     & LOCAL                               & 16                                   & LOCAL                               & 67                                   & LOCAL                               & 18                                    \\ 
\hline
\multirow{6}{*}{\begin{tabular}[c]{@{}c@{}}\textit{\textbf{High M}}\\\textit{\textbf{value}}\end{tabular}}                           & \multirow{2}{*}{SqueezNet: Conv 25} & RS                                  & \textbf{230}                         & OS                                  & 126                                  & WS                                  & 996                                   \\ 
\cline{3-8}
                                                                                                                                     &                                     & LOCAL                               & \textbf{6.6}                         & LOCAL                               & 16                                   & LOCAL                               & 31                                    \\ 
\cline{2-8}
                                                                                                                                     & \multirow{2}{*}{Resnet50: Conv 24}  & RS                                  & 74                                   & OS                                  & 116                                  & WS                                  & 42                                    \\ 
\cline{3-8}
                                                                                                                                     &                                     & LOCAL                               & 22                                   & LOCAL                               & 28                                   & LOCAL                               & 12                                    \\ 
\cline{2-8}
                                                                                                                                     & \multirow{2}{*}{VGG16: Conv 8}      & RS                                  & 351                                  & OS                                  & 98                                   & WS                                  & 411                                   \\ 
\cline{3-8}
                                                                                                                                     &                                     & LOCAL                               & 12                                   & LOCAL                               & 32                                   & LOCAL                               & 24                                    \\ 
\hline
\multirow{6}{*}{\begin{tabular}[c]{@{}c@{}}\textit{\textbf{High P}}\\\textit{\textbf{and Q}}\\\textit{\textbf{values}}\end{tabular}} & \multirow{2}{*}{SqueezNet: Conv1}   & RS                                  & 60                                   & OS                                  & 20                                   & WS                                  & \textbf{2238}                         \\ 
\cline{3-8}
                                                                                                                                     &                                     & LOCAL                               & 5.1                                  & LOCAL                               & 7                                    & LOCAL                               & 45                                    \\ 
\cline{2-8}
                                                                                                                                     & \multirow{2}{*}{Resnet50: Conv 1}   & RS                                  & 90                                   & OS                                  & 60                                   & WS                                  & 140                                   \\ 
\cline{3-8}
                                                                                                                                     &                                     & LOCAL                               & 6                                    & LOCAL                               & 13                                   & LOCAL                               & 23                                    \\ 
\cline{2-8}
                                                                                                                                     & \multirow{2}{*}{VGG16: Conv 1}      & RS                                  & 81                                   & OS                                  & 24                                   & WS                                  & 113                                   \\ 
\cline{3-8}
                                                                                                                                     &                                     & LOCAL                               & 6.6                                  & LOCAL                               & 6                                    & LOCAL                               & 17                                    \\
\hline
\end{tabular}
\end{table*} 

\section{Related Works}
DNN accelerators, like any other processing hardware, have two essential design points: components and data. Components are primary hardware resources, and data is the input of an accelerator. The focus of this paper is on the data side, more precisely, the data mapping design point. A basic concept of dataflow and mapping method is reusing data to reduce data movement between memory to PE-array, and PE to PE. Based on the category of \citep{chen2017using} the primary dataflow are, input \citep{chen2014diannao}, output \citep{du2015shidiannao}, \citep{gupta2015deep}, \citep{peemen2013memory}, weight \citep{NVDLA}, \citep{cavigelli2015origami}, \citep{chakradhar2010dynamically}, \citep{farabet2011neuflow}, \citep{park20154}, row stationary \citep{chen2016eyeriss} and no local reuse \citep{zhang2015optimizing}. 

Mapping strategy extended the idea of dataflow by reusing multiple parameters with specified ranges based on the shape of DNN and hardware resources of an accelerator. For instance, we can cite mRNA \citep{zhao2019mrna} for MAERI \citep{kwon2018maeri} and general methods including Marvel \citep{chatarasi2020marvel}, dMazeRunner \citep{dave2020dmazerunner} and interstellar \citep{yang2020interstellar} that improves energy consumption. 

Heuristic algorithms based on evolutionary \citep{kao2020gamma} or learning-based \citep{kao2020confuciux} methods have been proposed to improve energy consumption.

Several simulation frameworks have been developed to estimate primary parameters like latency and energy consumption. Timeloop \citep{parashar2019timeloop} employs Accelergy \citep{wu2019accelergy} to estimate energy consumption, mRNA \citep{zhao2019mrna} used the MAERI \citep{kwon2018maeri} energy model, and MAESTRO \citep{kwon2020maestro} employs an analytical model.

\section{Conclusion}
One of the primary aspects in the design of spatial DNN accelerators is data mapping. Evaluation results indicate how mapping affects energy efficiency.
But another critical issue that directly impacts compile time is mapping-time. This paper presents a low-complexity mapping algorithm called LOCAL, which finds an energy-efficient map rapidly. To reach a clear definition of the mapping strategy and the problem's scope, we formally represent the problem and goals. Simulation results indicate an immense reduction in the execution time of mapping, with a considerable improvement in energy consumption compared with other known dataflow mechanisms.
\section{Acknowledgment}
 This work was supported, in part, by Science Foundation Ireland under grant No. 13/RC/2094\_P2, co-funded under the European Regional Development Fund through the Southern \& Eastern Regional Operational Programme to Lero \footnote{\href{https://lero.ie/}{https://lero.ie/}} (Science Foundation Ireland Research Centre for Software), and, in part, this project has received funding from the European Union’s Horizon 2020 research and innovation programme under the Marie Skłodowska-Curie grant agreement No. 754489.

\bibliographystyle{unsrtnat}
\bibliography{references}  
\end{document}